\documentclass[pra,twocolumn,showpacs,preprintnumbers,amsmath,amssymb]{revtex4}

\usepackage{graphicx}
\usepackage{dcolumn}
\usepackage{bm}
\usepackage{mathrsfs}
\usepackage{epsfig}
\newcommand{\eq}[1]{Eq.~(\ref{#1})}
\newcommand{\fig}[1]{Fig.~\ref{#1}}
\newcommand{\be}[1]{\begin{equation}\label{#1}}
\newcommand{\ee}{\end{equation}}
\renewcommand{\vec}{\mathbf}

\begin{document}

\title{Double energy differential cross sections for the Coulomb four-body problem in a quasiclassical framework}

\author{Agapi Emmanouilidou}
\address{ITS, University of Oregon, Eugene, Oregon 97403-5203} \date{\today}
\begin{abstract}
In a quasiclassical framework, we formulate the double energy differential cross sections for the Coulomb
four-body problem.  We present results for the triple photoionization from the Li ground state at 220.5, 115, 50 and 3.8 eV excess energies. With the energy of one of the electrons kept fixed, the double energy differential cross sections at 220.5 and 115 excess energies are found to be of  ``U-shape" (unequal energy sharing), with those at 115 eV being in very good agreement with ab-initio
results. At 50 eV, it seems that a transition starts taking place to more equal energy sharing configurations. Close to threshold, at 3.8 eV excess energy, the equal energy sharing configurations are the dominant ones. \end{abstract}
\maketitle

\section{Introduction}
The single photon double ionization of the helium atom (Coulomb three-body problem) is by now well
stydied. There are several theoretical approaches and experimental results in good agreement with each other for total as well as for differential cross sections \cite{Briggs}. At the same time the single photon triple ionization of the lithium atom (Coulomb four-body problem) is still largely unexplored.
 
 Regarding the total triple photoionization cross section from the ground state of Li   
the few existing theoretical results \cite{Pindzola04,Pattard,Emmanouilidou1} are in good agreement with experiment \cite{Wehal98, Wehal00}. However, for differential cross sections of three escaping electrons the theoretical studies are very limited with no experimental results currently available. 
 The small number of theoretical studies and the current lack of experimental ones is partly due to the very small values of the differential cross sections. With the highest value of the total triple photoionization
 cross section for Li being a few barns, the differential ones are even smaller since they sample only part of the outgoing flux. However, there are recent experimental results for differential cross sections
 of three outgoing electrons for the electron-impact double ionization of helium. These experiments measure
 the angular distribution of two ionized electrons when the incoming electron is very fast \cite{Doern1}   and more recently measurements were obtained for incident electrons of much lower energy \cite{Doern2}.  
 The theoretical
work on differential cross sections for triple photoionization from the Li ground state include: a) A study of selection rules for different electron-momenta configurations in the three electron escape \cite{Malcherek}. In this work, various relative angular distributions of one of the outgoing electrons (for a given relative angle of the other two electrons) were presented using correlated 6 C (Coulomb) final state wavefunctions. b) There is also a very recent study of energy differential cross sections for photon energies of 280, 300 and 320 eV for the triple photoionization of Li using the non-perturbative time-dependent close coupling method \cite{Colgan}. c)
 Finally, in ref \cite{Emmanouilidou2} the angular correlation probability is evaluated in the framework of classical mechanics. The angular correlation depends only on the relative angle between any pair of ionized electrons in the three electron escape.

 In the current paper, we investigate double energy differential cross sections for the triple photoionization
 from the Li ground state. To our knowledge, it is the first time double energy differential cross sections are formulated classically. Energy differential cross sections for the complete break-up of the Li atom are much more difficult to interpret compared to those for the complete break-up of the He atom. 
 For two electron escape, for a given excess energy, there is only one energy differential cross section
 as a function of one electron's energy. For three electron escape, for a given excess energy, a different double energy differential cross section, as a function
 of one electron's energy, is obtained for each energy assigned to one of the electrons 
 (see below slices through differential cross sections). 
    Our results are obtained using a quasiclassical formulation which has been outlined elsewhere but we include it here as well, in section IIA, for completeness of the paper. 
   The classical nature of our investigation results in three distinguishable escaping electrons.  In section IIB, we outline how starting from the double energy differential cross sections of the three distinguishable electrons we obtain fully symmetrized ones.  Note that, in an ab-initio formulation the symmetry properties of the differential cross section are a natural outcome of using a fully antisymmetric wavefunction
to describe the three electron state. In section IIC, we discuss how the double energy differential cross sections are computed numerically. In section III, we present the double energy differential cross sections
for four different excess energies. For 115 eV 
our results are in very good agreement with the ab-initio double energy differential cross sections in ref\cite{Colgan} for a photon energy of 320 eV. Given that the ionization energy of the Li ground state is equal
to 203.5 eV, 320 eV photon energy corresponds to an excess energy close to 115 eV. 
 We find that  the ``U-shape"
is even more pronounced for an excess energy of 220.5 eV. At 50 eV, it seems that a transition starts taking place from a ``U-shape" to a more equal energy sharing configuration. We find that at 3.8 eV the slices through the double energy differential cross section have transitioned roughly to shapes where equal energy sharing is favored. 

\section{Quasiclassical formulation of single photon multiple ionization}  
     
\subsection{Initial phase space density and its time evolution for single photon triple ionization}
The construction  of the initial 
 phase space density
$\rho(\gamma)$ in our quasiclassical formulation of the
triple photoionization of Li has been detailed in \cite{Emmanouilidou1}, here we 
give only a brief summary.  We
formulate the triple photoionization process from the Li ground state
($1s^{2}2s$) as a two step process.
First, one electron absorbs the photon (photo-electron) at time 
$t=t_{\rm abs}=0$.  Through electronic correlations, the energy gets redistributed, 
resulting in three electrons escaping to the continuum.  Our formulation accounts for the second step.
  We first assume that the photon is absorbed by a $1s$-electron at the
nucleus ($\vec r_{1}=0$). This latter, approximation becomes exact in the
limit of high photon energy \cite{Kabir}.  The cross section for photon
absorption from a $1s$ orbital is much larger than from a $2s$ orbital
\cite{emsc+03}. Hence, we can safely assume that the photo-electron
is a $1s$ electron which significantly reduces the initial phase space
to be sampled.  Also, by virtue of their different character the
electrons become practically distinguishable and allow us to neglect
antisymmetrization of the initial state.  We denote the photo-electron
by 1, the other $1s$ electron by 2 and the $2s$ electron by 3.
 Following photon absorption, we model the initial phase space
distribution of the remaining two electrons, $1s$ and $2s$, by the
Wigner transform of the corresponding initial wavefunction $\psi({\bf
r}_{1}=0,{\bf r}_{2},{\bf r}_{3})$, where ${\bf r}_{i}$ are the
electron vectors starting at the nucleus.  We approximate the initial
wavefunction as a simple product of hydrogenic orbitals
$\phi^{\mathrm{Z}_{i}}_{i}(\vec r_{i})$ with effective charges
$Z_{i}$, to facilitate the Wigner transformation.  The $Z_{i}$ are
chosen so that they reproduce the known ionization potentials $I_{i}$, namely
for the 2s electron $Z_{3}=1.259$ ($I_{3}=0.198\,$a.u.) and for the 1s
electron $Z_{2}=2.358$ ($I_{2}=2.780\,$a.u.).  (We use atomic units
throughout the paper if not stated otherwise.)  The excess energy,
$E$, is given by $E=E_{\omega}-I$ with $E_{\omega}$ the photon energy
and $I=7.478$ a.u.\ the Li triple ionization threshold energy.
Given the above considerations, the initial phase space density
is given by
\begin{equation}
\label{eq:distribution}
\rho(\gamma) = \mathscr{N}
\delta(\vec{r}_1)\delta(E_{1}+I_{1}-\omega)\prod_{i=2,3}W_{\phi^{\mathrm{Z}_{i}}_{i}}
(\vec r_{i},\vec p_{i})\delta(E_{i}+I_{i})
\end{equation}
with normalization constant $\mathscr{N}$.
 
We next determine which fraction of $\rho(\gamma)$ leads to triple 
ionization, by following the phase space distribution in time.
  The evolution of a classical phase space density is determined by 
 the classical Liouville equation
  \begin{equation}
 \label{eq:Liouville}
 \frac{\partial \rho(\Gamma(t))}{\partial t}=\mathscr{L}_{\mathrm{cl}}\rho(\Gamma(t)),
\end{equation}
with the initial phase space values being 
\begin{equation}
 \label{eq:gamma}
\Gamma(0) \equiv \gamma\,,
\end{equation}
and $\mathscr{L}_{\mathrm{cl}}$ the classical Liouville operator
which is defined by the Poisson bracket \{H, \}, with H the
Hamiltonian of the system.  In our case H is the full Coulomb
four-body Hamiltonian.  In practice, \eq{eq:Liouville} amounts to
discretizing the initial phase space, assigning weights to each
discrete point $\gamma_{j}=(p_{j}(0),q_{j}(0))$ according to
$\rho(\gamma_{j})$, and evolving in time each initial condition
$\gamma_{j}$ with the Coulomb four-body Hamiltonian.  This amounts to
propagating electron trajectories using the classical equations of
motion (Classical Trajectory Monte Carlo method \cite{CTMC1,CTMC2}).  Regularized coordinates \cite{regularized} are used to
avoid problems with electron trajectories starting at the nucleus. 
  We
label as triple ionizing those trajectories with the energies off all three electrons being, asymptotically in time,
$E_{i}>0$, with $i=1,2,3$.

\subsection{Double differential probabilities}
Our goal is to formulate the double energy differential cross section ${ \rm d^{2}}{\sigma^{3+}}/{\rm d}{E_{\alpha}}{\rm d}{E_{\beta}}$. It should be such that when doubly integrating over it the total triple ionization cross section is recovered:
\begin{eqnarray}
\lefteqn{
\sigma^{3+}=\int_{0}^{E}{\rm d}{E_{\alpha}}\int_{0}^{E-E_{\alpha}}{\rm d}{E_{\beta}}\frac{ { \rm d^{2}}{\sigma^{3+}}}{{\rm d}{E_{\alpha}}{\rm d}{E_{\beta}}} }\nonumber\\
& &\equiv
\sigma_{abs}\int_{0}^{E}{\rm d}{E_{\alpha}}\int_{0}^{E-E_{\alpha}}{\rm d}{E_{\beta}}\frac{ { \rm d^{2}}{P^{3+}}}{{\rm d}{E_{\alpha}}{\rm d}{E_{\beta}}}.
\end{eqnarray}
$\sigma_{abs}$ is the total photoabsorption cross section at excess energy E, which
we take from experimental data for the Li ground state \cite{Wehlitz}. In what follows, we formulate
${ \rm d^{2}}{P^{3+}}/{\rm d}{E_{\alpha}}{\rm d}{E_{\beta}}$.

As it has already been mentioned in section IIA, by considering a product of hydrogenic orbitals as our initial state, we neglect antisymmetrization in the initial state. The three escaping
electrons are distinguishable resulting in three distinct double differential probabilities ${ \rm d^{2}}P^{3+}(E_{i},E_{j})/{\rm d}{E_{i}}{\rm d}{E_{j}}$ with the energy of the third electron being $E_{k}=E-E_{i}-E_{j}$ from conservation of energy and $i,j,k=1,2,3$, $i< j$ . We next symmetrize each
one of the above three double differential probabilities as follows:
 ${ \rm d^{2}}P^{3+}(E_{i},E_{j})/{\rm d}{E_{i}}{\rm d}{E_{j}}$ should be 
a) symmetric 
under exchange of the electron indices, that is, it should have the same value when $E_{i}\rightarrow E_{j}$ since electrons $i$ and $j$ are indistinguishable; b) for constant energy $E_{i}$ the double differential probability, which is now only a function of $E_{j}$, should
be symmetric with respect to $(E-E_{i})/2$ since electrons $j$ and $k$ are indistinguishable;
 for constant energy $E_{j}$ the double differential probability, which is now only a function of $E_{i}$, should
be symmetric with respect to $(E-E_{j})/2$ since electrons $i$ and $k$ are indistinguishable; 
 The double differential probability that satisfies the above properties is of the following form:

  \begin{align}
 &\frac{ { \rm d^{2}}{P^{3+}_{sym}(E_{i},E_{j})}}{{\rm d}{E_{i}}{\rm d}{E_{j}}}=\frac{1}{6}(\frac{ { \rm d^{2}}{P^{3+}(E_{i},E_{j})}}{{\rm d}{E_{i}}{\rm d}{E_{j}}}+\frac{ { \rm d^{2}}{P^{3+}(E_{j},E_{i})}}{{\rm d}{E_{i}}{\rm d}{E_{j}}} \nonumber\\
&+\frac{ { \rm d^{2}}{P^{3+}(E_{i},E-E_{i}-E_{j})}}{{\rm d}{E_{i}}{\rm d}{E_{j}}}+\frac{ { \rm d^{2}}{P^{3+}(E-E_{i}-E_{j},E_{i})}}{{\rm d}{E_{i}}{\rm d}{E_{j}}} \nonumber\\ 
&+\frac{ { \rm d^{2}}{P^{3+}(E_{j},E-E_{i}-E_{j})}}{{\rm d}{E_{i}}{\rm d}{E_{j}}}+\frac{ { \rm d^{2}}{P^{3+}(E-E_{i}-E_{j},E_{j})}}{{\rm d}{E_{i}}{\rm d}{E_{j}}}),
    \end{align}
 It now follows that the symmetrized double differential probability is given by:
\begin{align}
& \frac{{\rm d^{2}}P^{3+}(E_{\alpha},E_{\beta})}{{\rm d}{E_{\alpha}}{\rm d}{E_{\beta}}}=\sum_{i<j}
 \frac{1}{3}(\frac{ { \rm d^{2}}{P^{3+}_{sym}(E_{i},E_{j})}}{{\rm d}{E_{i}}{\rm d}{E_{j}}})
 \end{align}
where the normalization factor follows from
\begin{equation}
\int_{0}^{E}{\rm d}{E_{\alpha}}\int_{0}^{E-E_{\alpha}}{\rm d}{E_{\beta}}
\frac{ { \rm d^{2}}{P^{3+}(E_{\alpha},E_{\beta})}}{{\rm d}{E_{\alpha}}{\rm d}{E_{\beta}}}=P^{3+}.
\end{equation}

\subsection{Numerical evaluation of double differential probabilities}
To numerically evaluate the three ${ \rm d}P^{3+}(E_{i},E_{j})/{\rm d}{E_{i}}{\rm d}{E_{j}}$ we divide the energy
surface $[0,E][0,E]$ into $N^{2}$ equally sized square bins. We then find the triple ionized
trajectories which fall into the bins and add up their weights. Note that from conservation of energy the double energy differential probabilities map out a triangle, see section III. In the next section we present results
for 220.5, 115, 50 and 3.8 eV excess energy. The size of $N$ is chosen so that we have a small standard relative error for each square bin (the error is inversely proportional to the square root of the number of triple ionizing
events in a given square bin \cite{CTMC1}). For each of the
excess energies currently under investigation the number of triple ionizing trajectories used in our computations was no less than 8000. 

\section{Results}
\begin{figure}
\scalebox{0.35}{\includegraphics{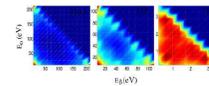}}
\caption{\label{fig4contour}Contour plots of the double energy differential cross section for excess energies 220.5, 115 and 3.8  eV. The triangle-like structure close to the $E_{\alpha}=E-E_{\beta}$ line is an artifact of our choice of square bins. }
\end{figure}

In this section we present results for excess energies of 220.5, 115, 50 and 3.8 eV. Our choice
of excess energies allows us to investigate the double differential cross sections for energies close to threshold (3.8 eV), close to the energy where the total cross section is maximum (50 eV) and for higher energies (115 eV and 220.5 eV). Note, that the 115 and 220.5 eV are in the energy range where our quasiclassical 
formulation is still valid (for very high excess energies one has to treat the problem quantum mechanically).
In \fig{fig4contour} we plot the double energy differential cross section for three excess energies, namely 220.5, 115 and 3.8 eV. The figure clearly illustrates that the double differential cross section has the symmetries discussed in section IIB. In addition, one can see that at large excess energies, 220.5 and 115 eV, the differential cross section has a bowl shape (with the bowl shape
at 220.5 eV being more pronounced). That is, for a given energy of one of the electrons the other two electrons share unequally the remaining energy, see below. On the other hand, at 3.8 eV 
for a given energy of one of the electrons the other two electrons tend to roughly equally share the remaining energy.

\begin{figure}
\scalebox{0.35}{\includegraphics{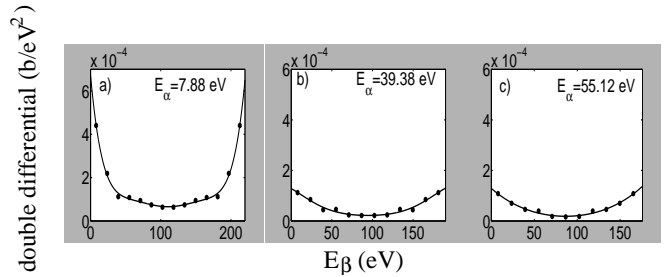}}
\caption{\label{fig1:220.5.5} Double energy differential cross section for $E=220.5$ eV. The cross section is shown
as a function of one electron's energy when the energy of one of the electrons is fixed to 7.88 eV, 39.38 eV and 55.12 eV. 
 Black circles are our data points, while the black line is a fit through these data points. The same holds for Figs \ref{fig2:115}, \ref{fig11:50}, \ref{fig3:3.8}. }
 \end{figure}

\begin{figure}
\scalebox{0.35}{\includegraphics{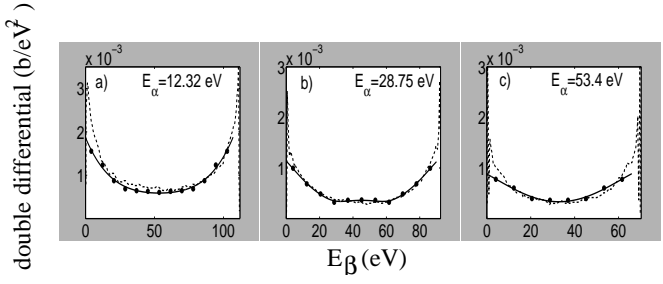}}
\caption{\label{fig2:115} Double energy differential cross section for $E=115$ eV. The cross section is shown 
as a function of one electron's energy when the energy of one of the electrons is fixed to 12.32 eV, 28.75 eV and 53.4 eV. Dashed black lines
are the data extracted from ref\cite{Colgan} for a 320 eV photon energy.
}
 \end{figure}

\begin{figure}
\scalebox{0.35}{\includegraphics{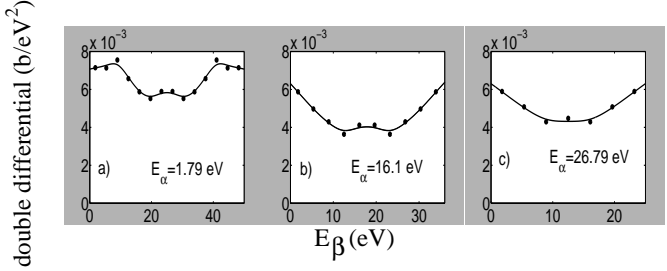}}
\caption{\label{fig11:50} Double energy differential cross section for $E=50$ eV. The cross section is shown
as a function of one electron's energy when the energy of one of the electrons is fixed to 1.79  eV, 16.1 eV and 26.79 eV. 
  }
 \end{figure}

\begin{figure}
\scalebox{0.35}{\includegraphics{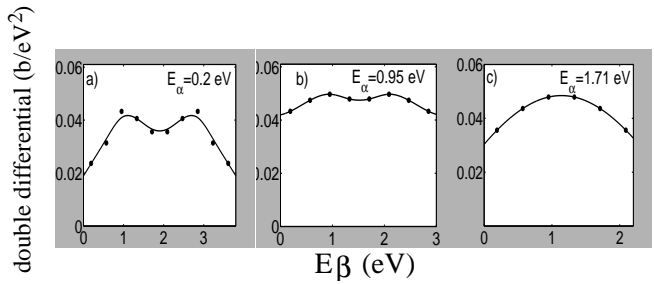}}
\caption{\label{fig3:3.8}Double energy differential cross section for $E=3.8$ eV. The cross section is shown
as a function of one electron's energy when the energy of one of the electrons is fixed to 0.2 eV, 0.95 eV and 1.71 eV.
 }
 \end{figure}

To gain more insight into the double differential cross sections, we consider in figures \ref{fig1:220.5.5}, \ref{fig2:115}, \ref{fig11:50} and \ref{fig3:3.8} slices through the double energy differential cross sections for each
of the four excess energies.
 That is, fixing the energy of one of the three escaping electrons, $E_{\alpha}$, we plot the double differential cross section as a function of the energy of another escaping electron, $E_{\beta}$. The energy of the third outgoing electron can be found from conservation of
energy. For all excess energies considered the energy surface $[0, E][0, E]$ is binned in $N^2$ squares 
with $N=14$ for 220.5, 115, 50 eV while $N=10$ for 3.8 eV. Our choice of N accurately accounts for the shapes of the slices through the
double differential cross section for the energies considered. 

In addition, for excess energies of 115 and 220.5 eV we see that the slices through the double differential cross section
have a ``U-shape''  which is more pronounced for the larger excess energy. The later is the well known characteristic shape of the single energy differential cross section for two outgoing electrons for high photon energies.
 In ref\cite{Colgan}, the slices through the
differential cross sections were all found to be of ``U-shape'' .  We next compare our quasicalssical results with the ab-initio results of ref \cite{Colgan} for 320 eV photon energy. The comparison
 is only an approximate one since our binning restricts the energies $E_{\alpha}$ we can consider. The value of $E_{\alpha}$ we use to compare is the one closer to the energy
 considered by the quantum calculations. Also note that the photon energy of 320 eV corresponds to a slightly higher excess energy than the 115 eV excess energy favoring a shape of slightly more unequal energy sharing than the ones considered in Figs 3a), 3b), 3c).
 We compare our results for 12.32, 28.75, 53.4 eV respectively with the results of ref\cite{Colgan} for
 10, 30, 50 eV (dashed line) in Figs 3a), 3b), 3c). Our data points are the black circles while
 the solid lines are a fit to these data points (the same holds for Figs \ref{fig1:220.5.5}, \ref{fig11:50}, \ref{fig3:3.8}).  We find that our results for 115 eV excess energy compare very well with those of ref\cite{Colgan}. The agreement is very good for intermediate values of $E_{\beta}$ while it is not as
 good close to the edges. However, close to the edges, as the authors in ref\cite{Colgan} point out, their results are less
 accurate due to some lack of convergence.  
 From Figs \ref{fig1:220.5.5},\ref{fig2:115} we see that for large excess energies of 115 and 220.5 eV, the smaller $E_{a}$ is the more pronounced the unequal energy sharing between the two electrons is. That means that for a very small energy of one electron a highly unequal energy sharing is favored among the other two. For a very large energy of one electron
 the unequal energy sharing of the other two is not as pronounced.

We next focus on the slices through the double differential cross sections for excess energies of 50 and 3.8 eV, with the later being very close to threshold, see \fig{fig11:50} and \fig{fig3:3.8}. One notices that 
at 50 eV for $E_{a}=1.79$ eV (see Fig. 4a)) the maximum of the differential cross section shifts to an energy $E_{\beta}\neq 0$ unlike the 115 and 220.5 eV excess energy where the maximum is at $E_{\beta}=0$. 
 In addition, one sees from \fig{fig11:50} that equal energy sharing among two of the electrons is
  favored for intermediate $E_{\beta}$ energies (hump in the middle of the $E_{\beta}$ energy range).
 It thus, seems, that at 50 eV excess energy a transition starts taking place from a highly unequal energy sharing at 220.5 and 115 eV to a more equal one.  At 3.8 eV, an energy very close to threshold, the transition to equal energy sharing configurations becomes even more pronounced. A comparison of Figs. \ref{fig11:50} and \ref{fig3:3.8} shows that the maximum of the differential cross section for 3.8 eV has shifted to energies
 $E_{\beta}$ higher than those for 50 eV and for $E_{\alpha}=1.71$ eV, see Fig. 5c), equal energy sharing is the favorable configuration. What is also quite interesting is that at  3.8 eV the slices
through the double differential cross sections as the energy $E_{\alpha}$ is increased, see Figs 5a), 5b), 5c), have shapes similar to those of the single energy differential cross sections of the two electron escape for decreasing excess energy close to threshold, see for example \cite{Rostb}.  

 It is clear that a thorough study of the slices through the double energy differentials for the three electron escape is needed for a large number of excess energies. Such a study can answer the question of how the transition
 takes place from a ``U-shape" at higher excess energies (220.5 and 115 eV) to more equal energy sharing for energies closer to the energy of the maximum  of the triple ionization cross section (50 eV) and finally to energies where equal energy sharing is the favorable configuration (3.8 eV). Understanding
 how the transition takes place will ultimately help us understand how the behavior of the three escaping electrons changes with decreasing excess energy.  
  To compare with two electron escape, let us point out that in the latter case a transition of the single energy differential cross section takes place from an unequal energy sharing to an equal sharing one as the photon energy
decreases. For the case of He double ionization, at 100 eV photon energy (close to the energy where the maximum of the double ionization cross section is)
the single energy differential cross section is flat. 100 eV is the critical energy where the transition from unequal to equal energy sharing starts taking place for two electron escape \cite{Proulx, Gailitis}.
 In addition, for the case of electron-hydrogen scattering \cite{Rostb}, as the excess energy is decreased (for energies close to threshold) it was found
 that the single differential cross sections have shapes similar to those shown in \fig{fig3:3.8}. What are the physical implications of this similarity
 for the three electron case remains to be seen.  
 
\section{Conlusions}
We have presented the first to our knowledge quasiclassical study of double energy differential cross sections for the triple ionization
of the lithium ground state. Our results for 115 eV are in good agreement with ab-initio calculations, indicating
that our simple initial state of a product wavefunction correctly captures the essential features of the triple ionization
process by single photon absorption from the Li ground state. We believe that our first results on how the transition
towards threshold takes place will be the impetus for future
studies of double energy differentials cross sections as the excess energy is reduced down to threshold. It could be the case that these detailed
studies for a large number of excess energies will allow a connection between the shapes of the double energy differential cross sections and the sequences
of collisions the three electrons follow to escape \cite{Emmanouilidou2} .

\begin{acknowledgments}
The author is indebted to Y. Smaragdakis, J. M. Rost and T. Pattard for discussions and for a critical reading of the manuscript.
\end{acknowledgments}

\clearpage

\begin{references}
\bibitem{Briggs}
J. S. Briggs and V. Schmidt J. Phys. B \textbf{33}, R1 (2000).
\bibitem{Pindzola04} J. Colgan, M. S. Pindzola, and F. Robicheaux  Phys.\ Rev.\ Lett. {\bf 93}, 053201 (2004) ;Phys. Rev. A \textbf{72}, 022727 (2005).
 \bibitem{Pattard} T. Pattard and J. Burgd\"orfer Phys. Rev A \textbf{63}, 020701(R) (2001).
\bibitem{Emmanouilidou1}
A. Emmanouilidou and J. M. Rost J. Phys. B \textbf{39}, L99 (2006).
\bibitem{Wehal98} R. Wehlitz, M. T. Huang, B. D. DePaola, J. C. Levin, I. A. Sellin, T. Nagata, J. W. Cooper, and Y. Azuma Phys. Rev. Lett. \textbf{81}, 1813 (1998).
\bibitem{Wehal00} R. Wehlitz, T. Pattard, M.-T. Huang, I. A. Sellin, J. Burgd\"orfer, and Y. Azuma, Phys. Rev. A, \textbf{61}, 030704(R) (2000).
\bibitem{Doern1}
A. Dorn, G. Sakhelashvili, C. H\"ohr, A. Kheifets, J. Lower, B. Najjari, C. D. Schr\"oter, R. Moshammer  and J. Ullrich \textit{Electron and Photon Ionization and Related Topics (Metz 2002)} ed L U Ancarani (Bristol: Institute of Physics Publishing) pp 41-51, 2002.
\bibitem{Doern2}
M. D\"urr, A. Dorn,J. Ullrich, S.P. Cao, A.S. Kheifets, J. R. G\"otz, J. S. Briggs, submitted (2006).
\bibitem{Malcherek}
A. W. Malcherek, J. M. Rost and J. S Briggs Phys. Rev. A \textbf{55}, R3979-82 (1997).
 \bibitem{Colgan}
 J. Colgan and M. S. Pindzola J. Phys. B \textbf{39}, 1879 (2006).
 \bibitem{Emmanouilidou2}
A. Emmanouilidou and J. M. Rost  J. Phys. B \textbf{39}, 4037 (2006); A. Emmanouilidou and J. M. Rost Phys. Rev. A accepted (2007). 
\bibitem{Kabir}
 P.K. Kabir and E. E. Salpeter Phys. Rev. {\bf 108}, 1256 (1950). 
 \bibitem{emsc+03} A. Emmanouilidou, T. Schneider and J. M. Rost 
J.~Phys.~B {\bf 36}, 2714 (2003). 
\bibitem{CTMC1}
R. Abrines and I. C. Percival Proc. Phys. Soc. London \textbf{88}, 861 (1966).  
\bibitem{CTMC2}
D. J. W. Hardie and R. E. Olson J. Phys. B \textbf{16}, 1983 (1983). 
\bibitem{regularized}
P. Kustaanheimo and E. Stiefel, J. Reine Angew. Math. \textbf{218}, 204 (1965).
\bibitem{Wehlitz}
R. Wehlitz 2004 private communication, see also \cite{Wehal00}
\bibitem{Rostb}
J. M. Rost Phys. Rev. Lett. \textbf{72}, 1998 (1994). 
\bibitem{Proulx}
D. Proulx and R. Shakeshaft Phys. Rev. A \textbf{48}, R875 (1993).
\bibitem{Gailitis}
M. Gailitis J. Phys. B \textbf{19}, L697 (1986).

\end{references}
\end{document}